\documentclass{elsart}
\usepackage{graphicx}% Include figure files
\usepackage[dvips]{epsfig}% Include figure files
\usepackage{dcolumn}% Align table columns on decimal point
\usepackage{bm}% bold math
\usepackage{amssymb}
\usepackage{amsmath}

\begin{document}
\begin{frontmatter}

\title{Particle Statistics and Population Dynamics}
\date{}

\author{Carlos Escudero\corauthref{cor}}
\corauth[cor]{Corresponding author, phone: (+34) 91-398-7126, fax: (+34) 91-398-6697.}
%\maketitle
\ead{cescudero@bec.uned.es}

\address{
Departamento de F\'{\i}sica Fundamental,
Universidad Nacional de Educaci\'on a Distancia, C/ Senda del Rey 9,
28040 Madrid, Spain}

\begin{abstract}
We study a master equation system modelling a population dynamics problem in a lattice. The problem is the calculation
of the minimum size of a refuge that can protect a population from hostile external conditions, the so called
critical patch size problem. We analize both cases in which the particles are considered fermions and bosons and show
using exact analitical methods that, while the Fermi-Dirac statistics leads to certain extinction for any refuge size,
the Bose-Eistein statistics allows survival even for the minimal refuge.
\end{abstract}

\end{frontmatter}

\section{Introduction}
\label{introduction}

Biological systems are known to be some of the most complex systems found in Nature. Lately, they have been
receiving great atention from the mathematical and physical sciences~\cite{murray}, that try to give
fundamental explanations to the phenomenology found in the field. One of the most important and changelling
problems in ecology is to understand the mechanisms that lead one population to extinction, interesting from
both the fundamental and applied point of view. Due to the complexity of the problem the most of the
approaches have been phenomelogical, like the use of stochastic differential equations~\cite{goel} or
reaction-diffusion equations~\cite{skellam}. Another consequence of this complexity is the lack of an
axiomatic theory describing this kind of systems. However, it has been noted that there is a deep connection between
ecological and reaction-diffusion processes~\cite{escudero};
a connection that may be exploted to get a more fundamental understanding
of ecology.

The problem under consideration is the so called critical patch size problem, already classic in the
mathematical literature. It was first considered by Skellam in his influencing article of
1951~\cite{skellam}, where he studied a population living in a finite refuge where the conditions are good
for life, but with hostile conditions outside. He showed that there is a critical size of the refuge such
that if the actual refuge is smaller than it the population get extincted. He used a reaction-diffusion equation
approach to that end, that considers a continuum population living in a continuum space. But since the
population is actually discrete and finite, we will use a different approach here, say, a master
equation approach, that considers a discrete population living in a discrete space. This
type of models have been
studied via Monte Carlo simulations and mean-field aproximations~\cite{weron1,weron2}, but here, instead, we will
perform exact analitical calculations. It is worth noting that Skellam's results have been improved recently in a
model that incorporates the internal fluctuations effects due to the discreteness of the population~\cite{escudero} (see also~\cite{doering} for a related problem on extinction).
This model, as well as Skellam's theory, deals with a continuum space and suppose every individual of the population punctual. The present model deals with a lattice, that may be interpreted as a correction to the infinitesimal size of the
individuals, while it represents the lost of the continuum space. This kind of analitical calculations are interesting
also as a guide for numerical simulations that take place, of course, on a lattice.

In this work we will get some insight into the extinction issue via two different models of reacting-diffusing particles:
in the first one the particles will obey the Fermi-Dirac statistics while in the second they will obey the Bose-Einstein
one. The biological interpretation is straightforward: the Fermi-Dirac statistics impose a maximum in the number of particles (individuals) per lattice site, i. e., a strict carrying capacity of the medium. In the case of Bose-Einstein statistics there is no such limit, even when some "crowded" realizations of the stochastic process could be highly unlikely.
It is easy to see that it is absolutely necessary to choose either statistics before studying a reaction-diffusion problem, be it numerical or theoretical.

At first look it seems that the statistics chosen has no relevance in this problem since the usual approach is to
consider the system near extinction~\cite{skellam,escudero,ludwig}. Indeed, the standard procedure is to linearize the
corresponding equation around the null population value, this is, to analize the low occupation number aproximation.
And since we are studying a system that seems to be controlled by its low occupation number state, this suggests that
the requirements on the occupation number (the particle statistics) will have no effect on the possible extinction event.
It is actually a question of theoretical interest if a system of reaction-diffusion bosons behave similarly to fermions in such systems dominated by a low occupation number state (see, for instance, the discussion in~\cite{cardy2}).
We will show that this is not the case. Actually, for the case of fermionic particles,
an arbitrarily well-adapted population will get extincted in finite time due
to a rare fluctuation event. This fact is very interesting also from the point of view of the stochastic
modelling, since rare fluctuations are some of worst understood issues that appear both in theory and
experiments, and whose understanding and control is very important for the pure science and applications
respectively~\cite{mc}. Also, the development of new techniques to study reaction-diffusion fermionic flows is a
very important subject that has proven itself very interesting for both points of view pure and
applied~\cite{mobilia1,mobilia2,mobilia3}.
In the case of bosonic particles
the situation changes completely and the population will be able to survive
even for the minimal refuge size. This implies that the importance of the statistics chosen for modeling the population
is complete.

In order to proof this statements we will use an interesting analogy with quantum mechanics, already common in the study of reaction-diffusion problems~\cite{cardy2,cardy} and that has proved itself useful when applied to biological systems~\cite{escudero,lopez}.

\section{The Fermionic Model}
\label{fermionic}

Our master equation will model a population of random walkers living in a finite lattice, that will be able
to reproduce, but will not neither compite for the nutrients nor die inside the refuge; death is only allowed
outside. This model can be thought as unreal, but we will study it because it overestimates the possibilites
of survival: if the population get extincted in this model, it will get extincted if we considered death
and/or competion. Every site of the lattice will be able to keep at most $N$ individuals, indicating the
finite resources of the medium to maintain a population. If we choose a low value of $N$, like $N=1$, we
are modelling a starving population, but because we can choose arbitrarily high values of $N$, we can model
arbitrarily well adapted populations.

Let us begin with the simplest case: A three sites one-dimensional chain undergoing the following series of
reactions. The first site (top to the left) allows diffusion to the right
\begin{equation}
A + \emptyset \to \emptyset + A
\end{equation}
at rate $D$, reproduction to the right
\begin{equation}
A + \emptyset \to A + A
\end{equation}
at rate $\alpha$, and death
\begin{equation}
A \to \emptyset
\end{equation}
at rate $\gamma$. The third site (top to the right) is symmetric to this one and allows death, reproduction
to the left, and diffusion to the left at the corresponding rates. The middle site allows diffusion and
reproduction to both the left and the right, but it is considered absolutely safe for the population, so no
death reaction is allowed. Additional reactions of death and competion can be added, but they will only
rend the extinction more likely. We will further consider a two-state chain: every site of the lattice
can be either empty or occupied by a single individual.
This system can be described via a master equation of the form:
\begin{equation}
\frac{dP_i(t)}{dt}=\sum_{j=1}^N [W(j \to i)P_j(t)-W(i \to j)P_i(t)],
\end{equation}
where $P_i(t)$ describes the probability of being in the state $\left| i \right>$ at time $t$, and
$W(j \to i)$ is the transition rate from the state $\left| j \right>$ to the state $\left| i \right>$.
It is worth pointing out that the algebraic properties of the master equation guarantee that its solution
consist of a linear combination of terms with a decaying exponential
time-dependence, and so will always show a stable approach to some steady
state~\cite{frensley,oppenheim}.

The
steady states of this equation are given by the condition $\frac{d\bar{P}_i(t)}{dt}=0$, and in our particular
case this condition reduces to the linear algebraic equation:
\begin{equation}
\label{steady}
\bar{\bar{M}} \cdot \vec{V} = \vec{0},
\end{equation}
where $\bar{\bar{M}}$ is the
$8 \times 8$~matrix
      \[ \left( \begin{array}{cccccccc}
     -2\gamma  & \alpha & 2\alpha & \alpha & 0 & 0 & 0 & 0 \\
     \gamma  & -T & D & 0 & \alpha & \alpha & 0 & 0 \\
     0 & D & -2T & D & 0 & 0 & 0 & 0 \\
     \gamma & 0 & D & -T & 0 & \alpha & \alpha & 0 \\
     0 & 0 & \gamma & 0 & -T & D & 0 & 0 \\
     0 & \gamma & 0 & \gamma & D & -2\alpha-2D & D & 0 \\
     0 & 0 & \gamma & 0 & 0 & D & -T & 0 \\
     0 & 0 & 0 & 0 & \gamma & 0 & \gamma & 0 \end{array} \right)\]
where $T=\alpha + D + \gamma$ and $\vec{V}$ is the vector
\begin{equation}
\left( \left| 111 \right>, \left| 110 \right>, \left| 101 \right>, \left| 011 \right>, \left| 100 \right>,
\left| 010 \right>, \left| 001 \right>, \left| 000 \right> \right)^t,
\end{equation}
that must fullfil the normalization condition
\begin{equation}
\label{normalization}
\left| 111 \right> + \left| 110 \right> + \left| 101 \right> + \left| 011 \right> + \left| 100 \right> +
\left| 010 \right> + \left| 001 \right> + \left| 000 \right> =1.
\end{equation}
In our notation $1$ stands for an occupied site and $0$ for an empty site, and $\left| ijk \right>$ for the
probability of being in such a state. The left hand side of Eq.(\ref{steady}) can be thought as a linear map;
this way our problem reduces to calculate its kernel. It is easy to see, by performing the matricial product
and solving the corresponding linear system, that, provided $\alpha>0$, $D>0$, and $\gamma>0$, the kernel
is generated by the linear span of the vector:
\begin{equation}
\left( 0,0,0,0,0,0,0,1 \right)^t.
\end{equation}
This is also the only element of the kernel that satifies the normalization condition
Eq.(\ref{normalization}), which means that it is the unique steady state
solution of the master equation. Physically,
this means that, no matter with which initial condition are we starting, the system will be in the state
$\left| 000 \right>$ in the infinite time limit. That is, \emph{extinction is certain} for long times.

This result is not surprising in view of the continuum space calculations of critical patch sizes, where one
can show that a patch shorter than the critical one leads to certain extinction, for a critical patch size
strictly greater than zero~\cite{skellam}.
In this discrete version of the problem, we have chosen the minimum possible
size of the refuge, so this result could be expected {\it a priori}. However, we would like to see what
happens in situations with different number of sites. This way we will see if it is possible to define
a critical patch number, that is, to determine what is the minimum number of sites being part of a refuge
that rends it effective to prevent an extinction in the infinite time limit.

Let us consider a one-dimensional two-state
finite lattice with $L+2$ sites. The first site (top to the left) allows
diffusion to the right, reproduction to the right and death at the rates defined below. The $(L+2)-th$ site
(top to the right) is defined as symmetric to the first one. The $L$ central sites allow reproduction and
diffusion both to the left and the right, but not death. Note that allowing only two sites to be out of the
refuge rends more difficult the extinction than if we consider more, since the individuals outside can only
diffuse inside. It is interesting to compare this with the continuum case, where it is considered an
infinite space out of the refuge at both boundaries.

In this case we can construct a master equation like in
the former one, and write the corresponding steady state condition generalizing Eq.(\ref{steady}). This will
be again a linear map given this time by a $2^{(L+2)} \times 2^{(L+2)}$~matrix.
Note that all the elements in the
$2^{(L+2)}-th$ column of this matrix are identically zero, reflecting the fact that
$\left| 0...0 \right>$ is an
absorbing state: once we arrive at this state, we will stay there forever. This is the mathematical
expression of the physical fact that no population can be created from nothing. This means that
$\left| 0...0 \right>$ always solves the master equation, something that should appear as obvious if we
realize that it can be chosen as initial condition, and therefore we will stay there for every $t>0$. For
$L=1$ we have also shown that the vacuum state is the unique steady state solution of the master equation,
implying that extinction is certain. We can establish therefore that the nonuniqueness of steady state
solutions of the master equation is a necessary condition to avoid extinction. This is equivalent to search
for situations in which the dimension of the kernel of the linear map has a dimension greater than or equal to
two. Provided that the determinant of the matrix is always zero (due to the fact that the smallest possible
kernel is one-dimensional), we are looking for situations in which the determinants of the minors of the
matrix are all zero. Fix $L>1$; in this case we have $2^{(L+2)} \times 2^{(L+2)}$ or less
algebraic equations of at most
$[(2^{(L+2)}-1) \times (2^{(L+2)}-1)]-th$ order,
which should be all zero in order to get a two-dimensional kernel. Suppose
that all this equations but one are identically zero, this way we arrive at a situation that is not
closer to extinction than the real one (we will consider the case of all equations identically zero below).
Fix $\alpha, D >0$; now we have a single algebraic equation in the variable $\gamma$ of at most
$[(2^{(L+2)}-1) \times (2^{(L+2)}-1)]-th$ order,
that has at most $(2^{(L+2)}-1) \times (2^{(L+2)}-1)$ solutions, as a consequence of
the fundamental theorem of algebra. Note that $\gamma = 0$ is a solution, since in this case we will arrive
certainly to the state $\left| 1...1 \right>$ in the infinite time limit, provided that the initial condition
is other than the vacuum state (in this case we can consider the null solution as ``unstable''). Suppose that
there are $N <(2^{(L+2)}-1) \times (2^{(L+2)}-1)$ solutions that are real and positive, and denote
$\gamma^*$ the smallest
of these solutions. For $\gamma = 0$ we have survival, as we have shown, and let us start now varying
continuously the value of $\gamma$ from zero to upper values. Because the set of solutions of an algebraic
equation is countable and finite, there is an open interval $(0,\gamma^*)$ of values of $\gamma$ that cause
certain extinction. That is, for values of $\gamma$ greater than zero but infinitely close to it, the system
is driven to extinction. As we vary continuously $\gamma$ we get certain extinction for an infinite number
of values till we arrive to $\gamma^*$, when we get again survival. This means that we get survival for
higher values of the death rate, while extinction is certain for lower values of it, something that is
absurd. We must necessarily conclude that there is not such $\gamma^*$, or what is the same, extinction is
certain for every $\gamma>0$. Note that in the case of more than one equation not identically zero, we can
reproduce the same argument defining $\gamma^*$ as the smallest common root of all the equations different
from zero.

Let us examine now the case in which all the equations are identically zero. Suppose that there is a value of
$L$, say $\bar{L}$, that fullfils this property. In this case, there should be at least another solution different from
the vacuum state, at which we should arrive if we start from the appropiate initial condition, no matter what
the values of the parameters are. Let fix $\gamma, D>0$, and $\alpha = 0$; in this case is easy to see that
we will get certain extinction in the infinite time limit for every initial condition, a contradiction. We
must necessarily conclude that there is not such $\bar{L}$, or what is the same, for every $L$ there is at
least one equation that is not identically zero. \emph{This means that, provided that
$\alpha, \gamma, D>0$, we get certain extinction in the limit $t \to \infty$}.

Generalizations of this problem are straightforward. Suppose that we have an hypercubic $d$-dimensional
lattice with $L$ sites per side. Suppose that we allow $N$ individuals per site, and diffusion and
reproduction to the first neighbours, and, in the case of $N>1$, also on-site reproduction. Suppose further
that we allow the death reaction to take place in just the boundary sites. A similar argument as the
above one will lead us again to the same conclusion: extinction is certain. Even in more complicated cases,
like when the occupation number depends on the lattice site $N=N(i)$, where this dependence may be
deterministic or stochastic (quenched disorder), we get the same result via an equivalent argument. We can
further claim that, because the probability of survival is identically zero in the steady state, all the
realizations of the stochastic process should reach the vacuum state in finite time.

The physical implications of this fact are very important. We can consider arbitrarily well adapted
populations by considering arbitrarily high (but bounded) values of $N$ and appropiate initial conditions.
We can further choose high values of the reproduction rate and low values of the death rate; no matter of
this the population will finally get extincted. This means that the dynamics of this system is dominated
by a rare fluctuation that always appears in finite time and kills the whole population.

\section{The Bosonic Model}
\label{bosonic}

In this section we will analize the same model but with bosonic particles for contrast. The Malthusian population under
consideration will undergo birth, $A \to A + A$, at rate $\sigma$, at every site, death, $A \to \emptyset$,
at rate $\gamma$, only at the boundary sites, and diffusing coupling between first neighbours.
To do the analysis we will need a very
different methodology. The master equation in this case reads
\begin{eqnarray}
\label{master1}
\nonumber
\frac{dP(\{n_i\};t)}{dt}=\sum_i \Big{(} D \sum_{\{e\}}[(n_e+1)P(...,n_i-1,n_e+1,...;t) \\
\nonumber
-n_iP(...,n_i,n_e,...;t)] \\
+\sigma[(n_i-1)P(...,n_i-1,...;t)-n_iP(...,n_i,...;t)] \Big{)},
\end{eqnarray}
where $\{e\}$ denotes the set of first neighbours of the site $i$ and
$i$ is not at the boundary. In the case $i$ is a boundary site the master equation is
\begin{eqnarray}
\label{master2}
\nonumber
\frac{dP(\{n_i\};t)}{dt}= \sum_i \Big{(} D \sum_{\{e\}}[(n_e+1)P(...,n_i-1,n_e+1,...;t) \\
\nonumber
-n_iP(...,n_i,n_e,...;t)]+ \\
\nonumber
\sigma[(n_i-1)P(...,n_i-1,...;t)-n_iP(...,n_i,...;t)]+ \\
\gamma[(n_i+1)P(...,n_i+1,...;t)-n_iP(...,n_i,...;t)] \Big{)}.
\end{eqnarray}
The analytical treatment of these equations comes as follows.
We can map this master equation description of the system into a quantum
field-theoretic problem. This connection was first proposed by Doi~\cite{doi} and
was deeply generalized in subsequent works~\cite{cardy}.
We can write this theory in terms of the second-quantized bosonic operators:
\begin{equation}
[a^\dag_i,a_j]=\delta_{ij}, \qquad [a_i,a_j]=0, \qquad [a^\dag_i,a^\dag_j]=0,
\qquad a_i \left| 0 \right>=0,
\end{equation}
whose effect is to create or to destroy particles at the corresponding lattice site:
\begin{equation}
a^\dag_i \left|...,n_i,...\right>=\left|...,n_i+1,...\right>,
\end{equation}
\begin{equation}
a_i \left|...,n_i,...\right>=n_i\left|...,n_i-1,...\right>,
\end{equation}
where we have defined the states as:
\begin{equation}
\left|\{n_i\}\right>=\prod_i (a_i^\dag)^{n_i} \left|0 \right>.
\end{equation}
Thus we can define the time-dependent state vector as:
\begin{equation}
\label{vector}
\left|\Phi(t) \right>=\sum_{\{n_i\}}P(\{n_i\};t)\left|\{n_i\} \right>,
\end{equation}
and claim that it obeys the imaginary time Schr\"{o}dinger equation
\begin{equation}
\label{schrodinger}
\frac{d}{dt}\left|\Phi(t)\right>=-H \left|\Phi(t) \right>,
\end{equation}
with the hamiltonian
\begin{equation}
\label{hamiltonian}
H=\sum_i \left(-D\sum_{\{e\}}a^\dag_i(a_e-a_i)
+\sigma [1-a^\dag_i]a^\dag_ia_i \right).
\end{equation}
Note that we recover Eq.(\ref{master1}) if we substitute Eq.(\ref{vector}) and Eq.(\ref{hamiltonian}) in
Eq.(\ref{schrodinger}). In the case of the linear chain, this second quantized theory is equivalent to the stochastic differential equation
\begin{equation}
da= D(a_{i+1}+a_{i-1}-2a_i)dt+\sigma a dt + \sqrt{2 \sigma a}dW(t),
\end{equation}
where $dW(t)$ denotes the increments of a Wiener process, and the correct interpretation of the equation is It\^{o}. This connection is explained in~\cite{cardy}, where it is also proved that the first moment of the stochastic process $a(t)$
is the first moment of the population. This implies
\begin{equation}
\frac{d\langle a_i \rangle}{dt}=D(\langle a_{i+1}\rangle +\langle a_{i-1}\rangle -2\langle a_i \rangle)+\sigma \langle a_i \rangle,
\end{equation}
in any site outside the boundary. If we repeat the calculation we get
\begin{equation}
\frac{d \langle a_i \rangle}{dt}=D(\langle a_e\rangle-\langle a_i \rangle)+(\sigma-\gamma) \langle a_i \rangle,
\end{equation}
when $i$ is at the boundary and $e$ denotes the first neighbour of $i$. This allows us to write the exact equation for the mean values of the population in the minimal chain
(L=3):
\begin{equation}
\frac{d\vec{a}}{dt}=\tilde{M} \cdot \vec{a},
\end{equation}
where $\vec{a}=(a_1,a_2,a_3)^t$ and $\tilde{M}$ is the $3 \times 3$~matrix
      \[ \left( \begin{array}{ccc}
     \sigma-\gamma-D & D & 0 \\
     D & \sigma-2D & D \\
     0 & D & \sigma-\gamma-D \end{array} \right)\]
We choose as initial condition $\vec{a}=(0,1,0)$, because is the minimal initial condition that keeps one particle under
the refuge. The solution can be calculated straightforwardly, but since the expresions are cumbersome we will only
report here the solution at the boundary, $a_b=a_1=a_3$,
\begin{eqnarray}
\nonumber
a_b(t)=-D\frac{\mathrm{exp}\left[\frac{1}{2}\left(-3D-\gamma-\sqrt{9D^2-2D\gamma+\gamma^2}+2\sigma\right)t\right]}
{\sqrt{9D^2-2D\gamma+\gamma^2}} \\
+D\frac{\mathrm{exp}\left[\frac{1}{2}\left(-3D-\gamma+\sqrt{9D^2-2D\gamma+\gamma^2}+2\sigma\right)t\right]}
{\sqrt{9D^2-2D\gamma+\gamma^2}}.
\end{eqnarray}
It is easy to see that the solution is increasing in time at some site if and only if it is increasing in time at
every site, thus it is enough to study the solution at one site, for instance, $a_b(t)$. And $a_b(t)$ is increasing
if and only if
\begin{equation}
\sigma \ge \frac{1}{2} \left(3D+\gamma-\sqrt{9D^2-2D\gamma+\gamma^2}\right).
\end{equation}
So we have seen that in the bosonic model the minimal chain allows survival for the appropiate parameter values.

\section{Conclusions}
\label{conclusions}

In this work, we have seen that the statistics of the reacting-diffusing particles modelling one ecological population
has a fundamental importance when we study the issue of extinction. It could be thought {\it a priori} that this would
not be the case since whenever we are close to extinction, due to the low number of surviving particles, restrictions
on the occupation number would not be important. We have shown that this is actually not the case, and that while
fermions are certainly driven to extinction for any refuge size, bosons can survive even in the minimal refuge.

We can also conclude that the risk of extinction can only be defined like a probability of
survival in a finite interval of time in the case of fermions. This definition has already been used in
biology~\cite{weron1,shaffer}, and it rules out the possibility of use those mathematical models
that suppose an absorbing boundary for enough high values of population. We have shown that
this supposition can not be assumed for enough long times, and we have put in a more rigorous footing
the definition first heuristically elucidated by the biologists.

But, in the case of bosons, the mathematical treatment for finite times is not necessary, because the infinite time
limit aproximation is enough to predict and separate surviving populations to those driven to extinction. Finally, as a remark, we would like to underline the methodological character of this work, that tries to clarify the mathematical
properties of the models commonly used to understand ecology, rather than give a realistic picture of the extinction
problem for one concrete population.

Very interesting mathematical problems
are still to be solved, for instance, to find the analytical expression of the time dependent solution of
the master equation in the case of the one-dimensional fermionic chain with $L+2$ sites.
This type of problem is usually
projected onto a spin chain problem expressed via Pauli operators, that turns out to be exactly integrable
in some cases~\cite{alcaraz}.

\section*{Acknowledgments}
This work has been partially supported by the Ministerio de Educaci\'{o}n y Cultura (Spain) through Grant
No. AP2001-2598, and by the Ministerio de Ciencia y Tecnolog\'{\i}a (Spain) through Project No. BFM2001-0291.

\end{document}